\documentclass[letterpaper,10pt]{article} 

\usepackage{opticameet3} 

\newcommand\authormark[1]{\textsuperscript{#1}}

\usepackage{amsmath,amssymb}
\usepackage[colorlinks=true,bookmarks=false,citecolor=blue,urlcolor=blue]{hyperref} 
\usepackage{subcaption}
\begin{document}

\title{Ethernet-over-OWC Using VCSELs: Transparent Gigabit Links with Low Latency and Robust Alignment Tolerance}




\vspace{-5mm}
\author{
    \footnotesize{Hossein Safi\textsuperscript{1*}, Isaac N. O. Osahon\textsuperscript{1}, Iman Tavakkolnia\textsuperscript{1}, and Harald Haas\textsuperscript{1}}
}
\address{\authormark{1} LiFi Research and Development Centre, Electrical Engineering Division, University of Cambridge, 9~J.~J.~Thomson Avenue, Cambridge CB3 0FA, UK.}

\email{\authormark{*}hs905@cam.ac.uk}

\vspace{-4mm}
\begin{abstract}
We demonstrate a fully bidirectional 1-Gb/s Ethernet-over-OWC link over a 1-m free-space path using a VCSEL–PIN pair and only commercially available components. The unamplified, transparent system achieves error-free operation, $<25$-ns latency, and centimetre-scale alignment tolerance.
\end{abstract}

\vspace{0mm}
\section{Introduction}

Optical wireless communication (OWC), which uses free-space light to complement millimeter-wave and sub-terahertz bands, offers high capacity, immunity to electromagnetic interference, and energy-efficient operation~\cite{LiFi-2}. These advantages make OWC a compelling solution for short-range, high-throughput links in environments where conventional radio frequency (RF) technologies face bandwidth or interference constraints.

At the physical layer, vertical-cavity surface-emitting lasers (VCSELs) are particularly attractive for realizing compact and energy-efficient OWC transmitters. Their low threshold currents, high modulation bandwidths, and wafer-scale manufacturability make them ideal for scalable and cost-effective deployment in emerging 6G systems~\cite{VCSEL-1}. In this work, we employ a custom-developed 980-nm VCSEL specifically designed to deliver high-speed, low-power optical transmission suitable for real-time OWC links. The device exhibits stable single-mode operation, low threshold current, and multi-gigahertz modulation bandwidth, offering a compact and energy-efficient light source well suited for short-range, high-throughput optical wireless systems.

To transition from device-level demonstrations to practical networking systems, OWC must integrate seamlessly with existing digital infrastructure. Ethernet, standardized under IEEE~802.3, spans data rates from 10~Mb/s (10BASE-T) to 400~Gb/s and beyond, forming the backbone of enterprise, industrial, and automotive networks \cite{EthernetIEEE2022}. A transparent Ethernet-over-OWC link acts as a direct wireless substitute for a physical Ethernet cable, maintaining native Ethernet protocols without conversion. This enables seamless optical connectivity in dynamic, cable-free environments like smart factories, data centers, and automotive systems.

Previous Ethernet-based OWC demonstrations were typically limited to sub-100-Mb/s rates and relied on amplified or complex optical front ends such as avalanche photodiodes or erbium-doped fiber amplifiers ~\cite{Ethernet-1,Ethernet-2}. Others were restricted to unidirectional transmission, lacking the full networking capability required for plug-and-play operation~\cite{Ethernet-3}. Reported gigabit-level demonstrations~\cite{Ethernet-4} have often employed fiber-coupled sources, which are highly sensitive to vibration and sub-millimetre misalignment, limiting practical deployment.

In this work, we overcome these limitations by demonstrating the first bidirectional, real-time, and fully transparent Ethernet-over-OWC link implemented entirely with off-the-shelf components and unamplified optical front ends. The transmitter employs a custom-developed 980-nm VCSEL optimized for stable, low-power, and high-speed operation, enabling reliable free-space optical transmission without amplification. The system functions as a true full-duplex Ethernet bridge over a 1-m free-space channel, achieving throughput above 900~Mb/s, sub-25-ns physical-layer latency, and centimetre-scale alignment tolerance. This combination of simplicity, real-time operation, and robustness establishes a new benchmark for practical OWC deployment as a direct wireless extension of wired Ethernet.

\vspace{-2mm}

\section{Experimental Setup}
\begin{figure}[hbt!]
	\centering
	\includegraphics[height=5.2cm]{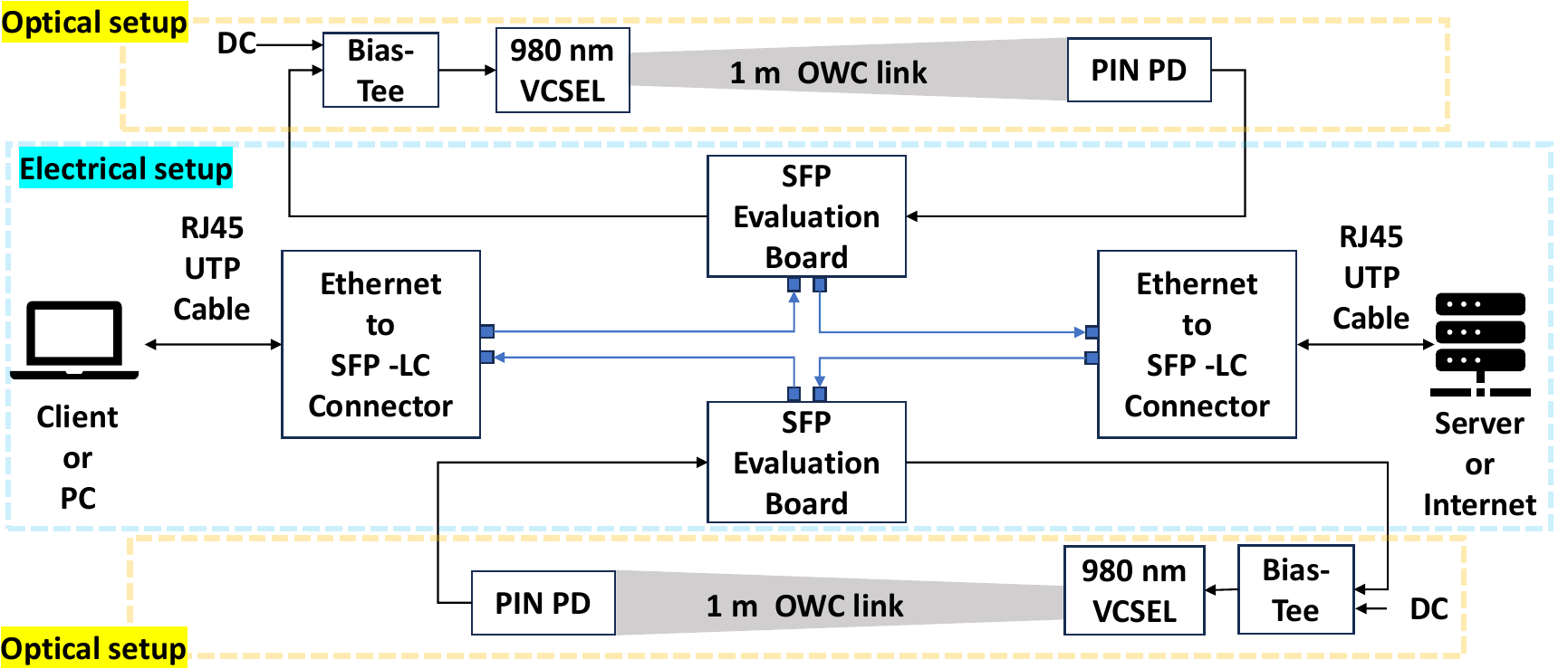}
	\caption{Schematic of the experimental setup for a bidirectional 1-Gb/s Ethernet-over-OWC link. Standard Ethernet traffic is converted to optical via media converters and SFP modules, modulated by a VCSEL transmitter, transmitted across a 1-m optical wireless channel, and detected by a PIN photodiode before being reconverted to Ethernet. The reverse path implements full-duplex operation.} 
	\label{Schematic}
\end{figure}
\begin{figure}[hbt!]
	\centering
	\includegraphics[width=\textwidth]{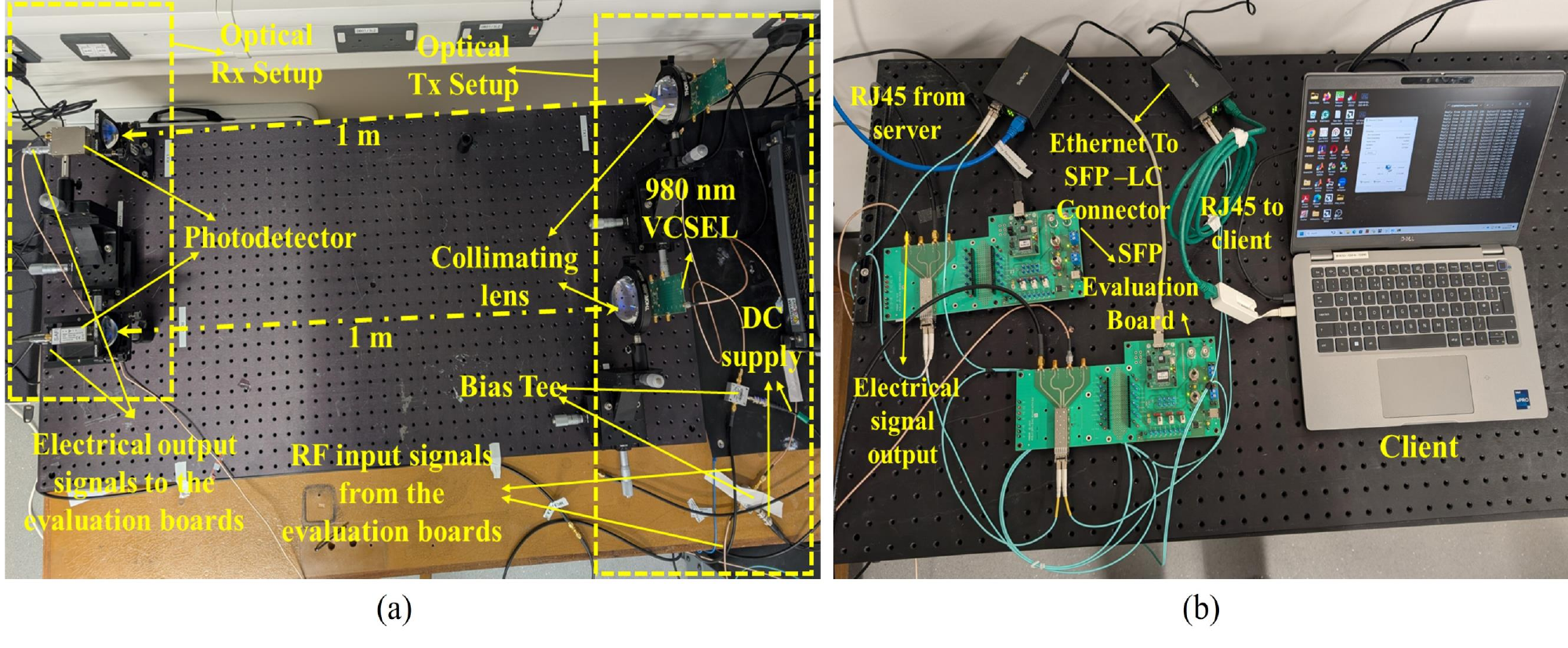}
	\caption{Photographs of the experimental Ethernet-over-OWC link setup corresponding to Fig. \ref{Schematic}. (a) Optical setup showing the Tx and Rx optical subsystems. (b) Electrical setup comprising the server, media converters, and SFP evaluation boards, collectively implementing a fully transparent bidirectional Ethernet bridge.} 
	\label{Schematic2}
    \vspace{-8mm}
\end{figure}
The experimental system was designed to establish a fully transparent, bidirectional Ethernet link over an optical wireless channel using only commercially available components. The overall system architecture is illustrated in Fig.~\ref{Schematic}, and photographs of the physical setup are shown in Fig.~\ref{Schematic2}.

On the transmit side, a server generated standard Ethernet traffic, which was delivered via copper (RJ45) to an Ethernet-to-SFP media converter (MC220L, TP-Link). The converter provided a 1000BASE-SX optical output through an LC connector, coupled to a hot-pluggable small form-factor pluggable (SFP) transceiver module. The optical signal was then routed to an SFP evaluation board (Finisar FDB-1032-SFP), where it was converted into a high-speed electrical signal used to drive the optical wireless transmitter.
The transmitter employed a 980-nm VCSEL. The electrical output from the evaluation board was combined with a DC bias to directly modulate the VCSEL, eliminating the need for amplification or pre-emphasis circuitry. The emitted beam was collimated using a single aspherical lens and launched across a 1-m free-space optical channel.

At the receiver, the optical signal was detected by a FEMTO HSPRX-I-1G4-SI-FS silicon PIN photodiode with a 1.4-GHz bandwidth. The photodiode converted the received optical signal into the electrical domain, which was fed into a second SFP evaluation board. The recovered electrical signal was re-encoded into the optical domain via the SFP transceiver and subsequently converted back to copper Ethernet using a second media converter before being delivered to the client laptop. The reverse path was implemented symmetrically to realize a true bidirectional Ethernet bridge, enabling full-duplex operation. The overall system thus functioned as a plug-and-play optical wireless replacement for a physical Ethernet cable, requiring no protocol conversion or software modifications.

\section{Results and Discussion}
For performance monitoring, the output of the evaluation board was also connected to a Keysight UXR0104B real-time oscilloscope to capture eye diagrams and evaluate signal quality. Latency measurements were performed using a Keysight P9370A vector network analyzer (VNA), while throughput and real traffic performance were validated using standard online speed testing tools (speedtest.net).

\begin{figure}[hbt!]
    \centering
    \begin{subfigure}[b]{0.40\linewidth}
        \centering
        \includegraphics[width=\linewidth]{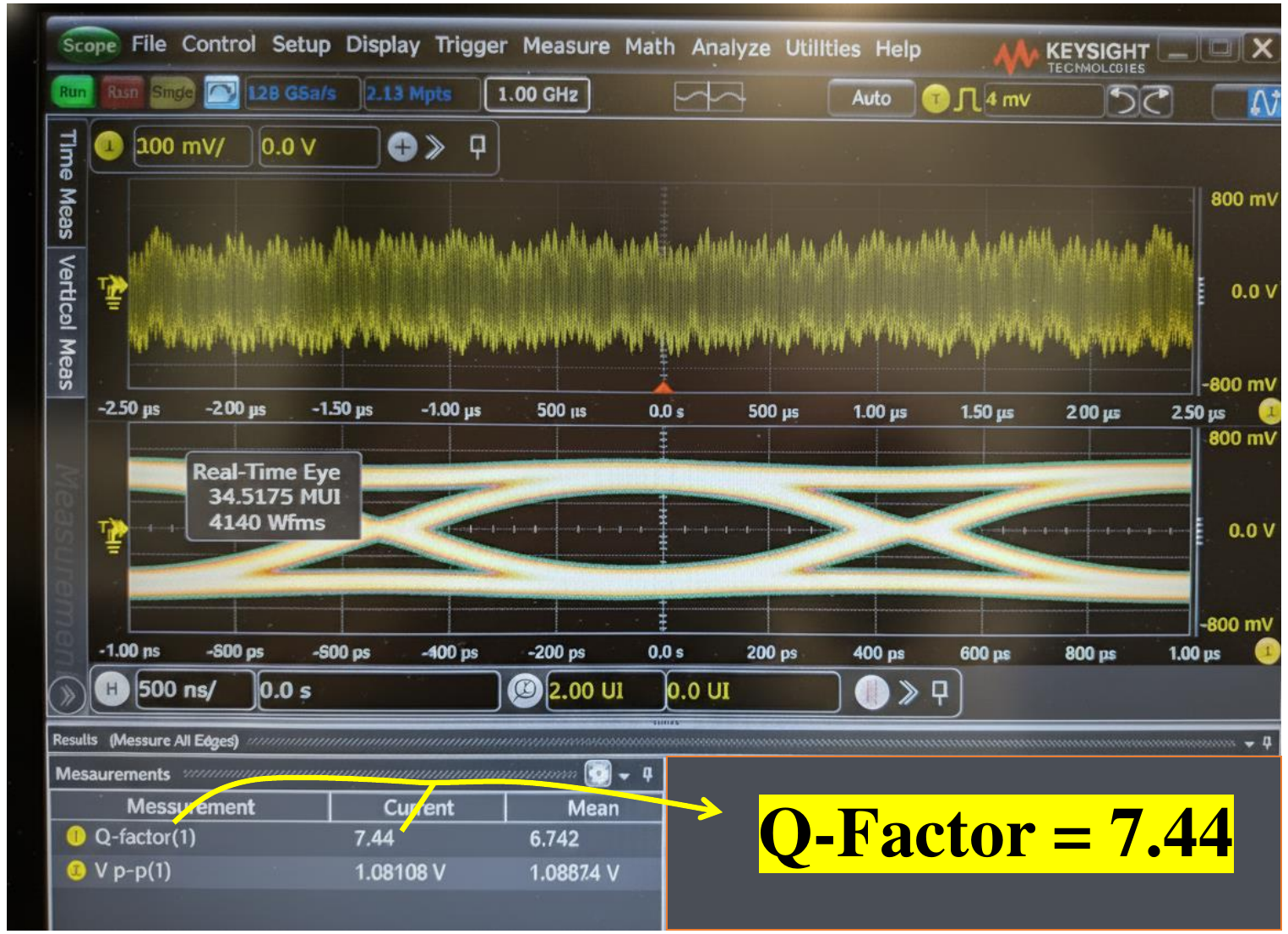}
        \caption{Eye diagram of the 1-Gb/s link showing a clear eye opening and $Q=7.44$.}
        \label{fig:eye_diagram}
    \end{subfigure}
    \hfill
    \begin{subfigure}[b]{0.50\linewidth}
        \centering
        \includegraphics[width=\linewidth]{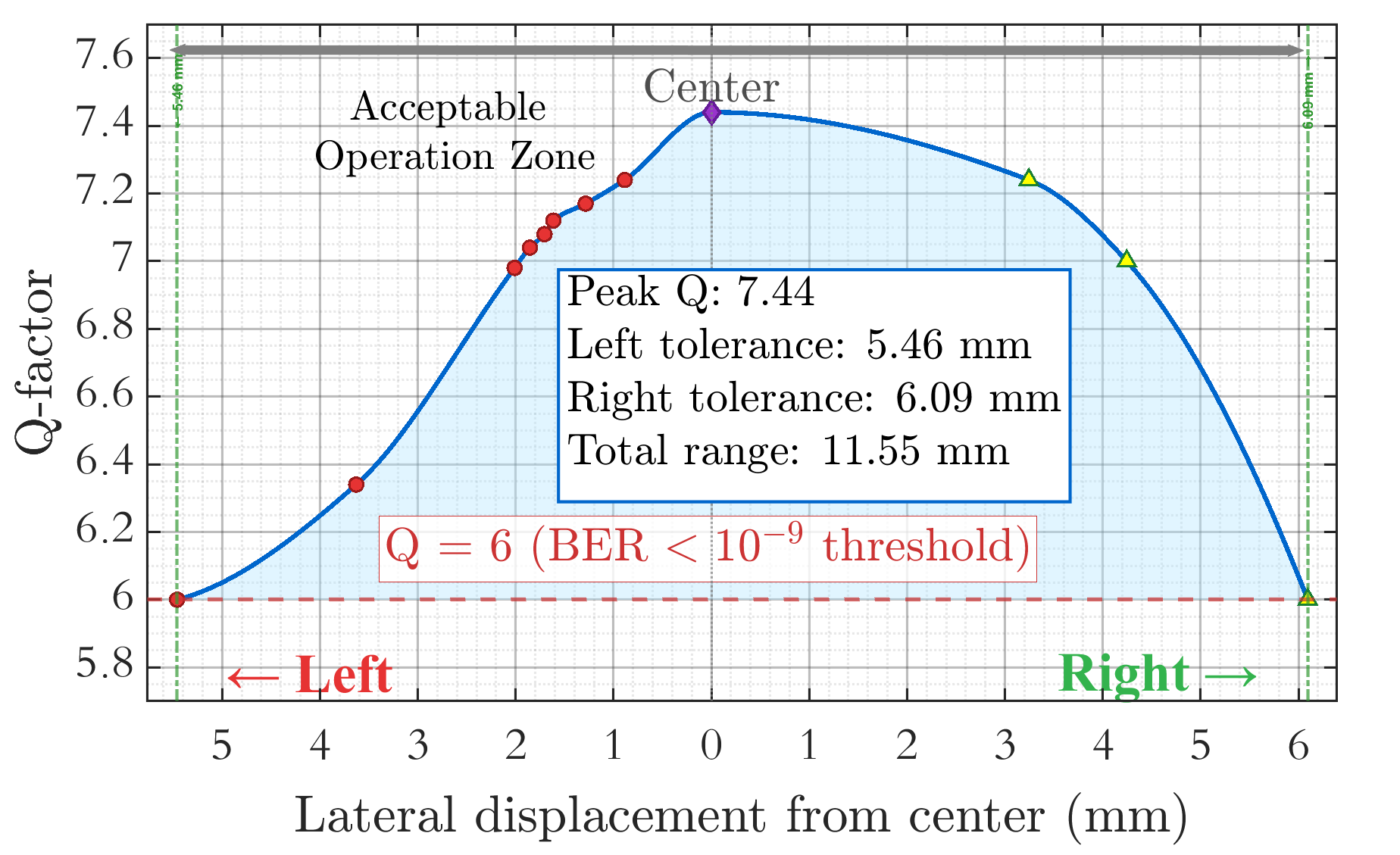}
        \caption{Measured Q-factor versus receiver lateral displacement.}
        \label{fig:qfactor_plot}
    \end{subfigure}
    \caption{(a) Measured eye diagram and (b) Q-factor performance of the Ethernet-over-OWC link. }
    \label{Measurement}
    \vspace{-4mm}
\end{figure}
\subsection{Signal Quality and Throughput Performance}
Fig.~\ref{Measurement} shows the measured Q-factor and eye diagram for the 1-Gb/s link. At optimal alignment, a Q-factor of 7.44 was obtained. Using the Gaussian approximation, the BER can be estimated from the Q-factor as $\mathrm{BER} = \tfrac{1}{2}\,\mathrm{erfc}\!\left(\tfrac{Q}{\sqrt{2}}\right)$, which yields a BER of approximately $10^{-13}$, well within the Ethernet requirement of $10^{-12}$~\cite{EthernetIEEE2022,AgrawalFiberOptics}. The wide eye opening confirms that the optical wireless channel introduces negligible distortion at this data rate. Real traffic was also tested using speedtest.net between a server and a client bridged by the optical wireless link. Download and upload rates of 813.95 Mb/s and 929.53 Mb/s, respectively, were achieved. These results approach the theoretical maximum of Gigabit Ethernet once protocol overheads are accounted for, confirming that the OWC links function as transparent extensions of the wired network.

\subsection{Misalignment Tolerance}

As shown in Fig. \ref{fig:qfactor_plot}, we evaluated the link’s tolerance to receiver displacement by gradually shifting it left and right. The Q-factor was 7.44 at the central position, dropping to 6 when moved 5.46~mm left or 6.09~mm right. This gives a total alignment tolerance of approximately 1.155~cm while maintaining a BER below $<10^{-9}$. This centimetre-scale tolerance is far more relaxed than typical fiber-coupled free-space links, which often require sub-millimetre alignment, highlighting the setup’s suitability for practical deployment.

\subsection{Latency}
End-to-end latency was measured using a vector network analyzer (VNA) by extracting the group delay between a back-to-back reference and the transmitted signal. Over the 1-m free-space path, the optical wireless link introduced less than 25~ns total delay, including $\sim$3.3~ns propagation, confirming near-transparent performance suitable for latency-sensitive applications.

%
\vspace{-2mm}
\section{Conclusion}
We demonstrated a bidirectional, real-time, and fully transparent Ethernet-over-OWC link using only off-the-shelf components and a single unamplified VCSEL–PIN photodiode pair per direction. The 1-Gb/s full-duplex link over a 1-m free-space channel achieved error-free, real-time transmission with throughput exceeding 900~Mb/s (Q~$>$~7), sub-25-ns physical-layer latency, and robust performance under lateral misalignments up to 1.16~cm.
\newline
\small{This work is a contribution by \textit{Project REASON}, a UK Government-funded project under the Future Open Networks Research Challenge (FONRC) sponsored by the Department of Science, Innovation and Technology (DSIT). }




\vspace{-3mm}
\begin{thebibliography}{99} 



\bibitem{LiFi-2}
A. Krishnamoorthy \textit{et al.}, ``Optical Wireless Communications: Enabling the Next-Generation Network of Networks,'' in IEEE Veh. Technol. Mag., vol. 20, no. 2, pp. 20-39, June 2025.

\bibitem{VCSEL-1} 
Yi Liu \emph{et al.}, ``High-Capacity Optical Wireless VCSEL Array 	Transmitter With Uniform Coverage,'' \textit{Free-Space Laser Communications XXXV}, vol. 12413,  p. 124130J, SPIE, 2023.

\bibitem{EthernetIEEE2022}
IEEE Std 802.3-2022, ``IEEE Standard for Ethernet,'' in \textit{IEEE Std 802.3-2022 (Revision of IEEE Std 802.3-2018)}, IEEE, Sept., 2022.


\bibitem{Ethernet-1}
G. Cossu \textit{et al.}, ``Sea-Trial of Optical Ethernet Modems for Underwater Wireless Communications,'' in \textit{J. Light. Technol.}, vol. 36, no. 23, pp. 5371-5380, 1 Dec. 2018.

\bibitem{Ethernet-2}
A. Shrestha \textit{et al.}, ``Inter-Island Demonstration of an FSO High Speed Laser Ethernet Transceiver for Telerobotic Space-Surface Control,'' \textit{42nd European Conf. on Optic. Commun. (ECOC)}, Dusseldorf, Germany, 2016.

\bibitem{Ethernet-3} 
G. Cossu \emph{et al.}, ``Real-Time Gigabit-Ethernet Transmission over Optical Wireless Using Off-the-Shelf Components,'' \textit{43rd European Conf. on Optic. Commun. (ECOC)}, Gothenburg, Sweden, 2017.

\bibitem{Ethernet-4}
M. M. Abadi \textit{et al.}, ``Implementation and Evaluation of a Gigabit Ethernet FSO Link for `The Last Metre and Last Mile Access Network',''  \textit{IEE Int. Conf. Commun. Workshops (ICC Workshops)}, Shanghai, China, 2019.

\bibitem{AgrawalFiberOptics}
G. P. Agrawal, \textit{Fiber-Optic Communication Systems}. Hoboken, NJ,
USA: Wiley, 2012.

\end{thebibliography}
\end{document}